\documentclass[twocolumn]{aastex7}

\usepackage{amsmath}

\begin{document}

\title{The BINGO/ABDUS Project: \\
Forecast for cosmological parameters from a mock Fast Radio Bursts survey}

\author{Xue Zhang}
\affiliation{Center for Gravitation and Cosmology, College of Physical Science and Technology, Yangzhou University, Yangzhou 225009, China}
\email[show]{zhangxue@yzu.edu.cn}

\author{Yu Sang}
\affiliation{Center for Gravitation and Cosmology, College of Physical Science and Technology, Yangzhou University, Yangzhou 225009, China}
\email[show]{sangyu@yzu.edu.cn}

\author{Gabriel A. Hoerning}
\affiliation{Jodrell Bank Centre for Astrophysics, Department of Physics \& Astronomy, The University of Manchester, Oxford Road, Manchester, M13 9PL, U.K. }
\affiliation{Instituto de F\'{i}sica, Universidade de S\~ao Paulo, R. do Mat\~ao, 1371 - Butant\~a, 05508-09 - S\~ao Paulo, SP, Brazil}
\email[show]{gabrielamancio.hoerning@postgrad.manchester.ac.uk}

\author{Filipe B. Abdalla}
\affiliation{Department of Astronomy, School of Physical Sciences, University of Science and Technology of China, Hefei, Anhui 230026, China}
\affiliation{CAS Key Laboratory for Research in Galaxies and Cosmology, University of Science and Technology of China, Hefei, Anhui 230026, China}
\affiliation{School of Astronomy and Space Science, University of Science and Technology of China, Hefei, Anhui 230026, China}
\affiliation{Instituto de F\'{i}sica, Universidade de S\~ao Paulo, R. do Mat\~ao, 1371 - Butant\~a, 05508-09 - S\~ao Paulo, SP, Brazil}
\affiliation{Department of Physics and Electronics, Rhodes University, PO Box 94, Grahamstown, 6140, South Africa}
\email[show]{filipe.abdalla@gmail.com} 

\author{Elcio Abdalla}
\affiliation{Instituto de F\'{i}sica, Universidade de S\~ao Paulo, R. do Mat\~ao, 1371 - Butant\~a, 05508-09 - S\~ao Paulo, SP, Brazil}
\affiliation{Universidade Estadual da Paraiba, Rua Bara\'{u}nas, 351, Bairro Universitário, Campina Grande-PB}
\affiliation{Departamento de F\'{i}sica, Centro de Ci\^{e}ncias Exatas e da Natureza - CCEN, Universidade Federal da Para\'{i}ba - UFPB || CEP 58059-970 -- João Pessoa - PB}
\email{}

\author{Amilcar Queiroz}
\affiliation{Unidade Acad\^emica de F\'{i}sica, Universidade Federal de Campina Grande, R. Apr\'{i}gio Veloso,  Bodocong\'o, 58429-900 - Campina Grande, PB, Brazil}
\email{}
    
\author{Andr\'{e} A. Costa}
\affiliation{Center for Gravitation and Cosmology, College of Physical Science and Technology, Yangzhou University, Yangzhou 225009, China}
\affiliation{College of Physics, Nanjing University of Aeronautics and Astronautics, Nanjing 211106, China}
\affiliation{Instituto Federal de Educa\c{c}\~{a}o, Ci\^{e}ncia e Tecnologia de Minas Gerais, Campus Itabirito, Itabirito 35450-000, Brazil}
\affiliation{Departamento de F\'{i}sica, Universidade Federal da Para\'{i}ba, Caixa Postal 5008, Jo\~{a}o Pessoa 58051-900, Para\'{i}ba, Brazil}
\email{}

\author{Ricardo G. Landim}
\affiliation{Technische Universit\"at M\"unchen, Physik-Department T70, James-Franck-Strasse 1, 85748, Garching, Germany}
\affiliation{Institute of Cosmology and Gravitation, University of Portsmouth, Dennis Sciama Building, Portsmouth PO1 3FX, United Kingdom}
\email{}

\author{Chang Feng} 
\affiliation{Department of Astronomy, School of Physical Sciences, University of Science and Technology of China, Hefei, Anhui 230026, China}
\affiliation{CAS Key Laboratory for Research in Galaxies and Cosmology, University of Science and Technology of China, Hefei, Anhui 230026, China}
\affiliation{School of Astronomy and Space Science, University of Science and Technology of China, Hefei, Anhui 230026, China}
\email{}

\author{Bin Wang} 
\affiliation{Center for Gravitation and Cosmology, College of Physical Science and Technology, Yangzhou University, Yangzhou 225009, China}
\affiliation{School of Aeronautics and Astronautics, Shanghai Jiao Tong University, Shanghai 200240, China}
\email{}

\author{Marcelo V. dos Santos} 
\affiliation{Fundação Coppetec, Rua Moniz de Aragão, 360 - Bloco 1 - Cidade Universitária da Universidade Federal do Rio de Janeiro (UFRJ), CEP 21941-594, Rio de Janeiro, RJ, Brazil}
\affiliation{Núcleo de Pesquisa Aplicada à Meios Geológicos-Geotécnicos, Av. Athos da Silveira Ramos, 149 - Bloco I, 203 - Cidade Universitária da Universidade Federal do Rio de Janeiro (UFRJ), Rio de Janeiro - RJ, 21941-909}
\email{}

\author{Thyrso Villela} 
\affiliation{Instituto Nacional de Pesquisas Espaciais - INPE, Av. dos Astronautas 1758, Jardim da Granja, S\~ao Jos\'e dos Campos, SP, Brazil}
\affiliation{Centro de Gest\~ao e Estudos Estrat\'egicos - CGEE, SCS, Qd 9, Lote C, Torre C S/N, Salas 401 - 405, 70308-200 - Bras\'ilia, DF, Brazil}
\affiliation{Instituto de F\'{i}sica, Universidade de Bras\'{i}lia, Campus Universit\'ario Darcy Ribeiro, 70910-900 - Bras\'{i}lia, DF, Brazil}
\email{}

\author{Carlos A. Wuensche}
\affiliation{Instituto Nacional de Pesquisas Espaciais - INPE, Av. dos Astronautas 1758, Jardim da Granja, S\~ao Jos\'e dos Campos, SP, Brazil}
\email{}

\author{Jiajun Zhang}
\affiliation{Shanghai Astronomical Observatory, Chinese Academy of Sciences, Shanghai 200030, China} 
\email{}

\author{Edmar Gurj\~ao}
\affiliation{Unidade Acadêmica de Engenharia Elétrica, Universidade Federal de Campina Grande, R. Aprígio Veloso, Bodocongó, 58429-900 Campina Grande, PB, Brazil}
\email{}

\author{Alessandro Marins} 
\affiliation{Department of Astronomy, School of Physical Sciences, University of Science and Technology of China, Hefei, Anhui 230026, China}
\email{}

\author{Alexandre Serres}
\affiliation{Unidade Acadêmica de Engenharia Elétrica, Universidade Federal de Campina Grande, R. Aprígio Veloso, Bodocongó, 58429-900 Campina Grande, PB, Brazil}
\email{}

\author{Linfeng Xiao}
\affiliation{Department of Physics and Astronomy, Sejong University, Seoul, 143-747, Korea}
\email{}



\begin{abstract}

There are various surveys that will provide excellent data to search for and localize Fast Radio Bursts (FRBs). The BINGO project will be one such survey, and this collaboration has already estimated a FRB detection rate that the project will yield. We present a forecast of the future constraints on our current cosmological model that the BINGO FRB detections and localizations will have when added to other current cosmological datasets.
We quantify the dispersion measure (DM) as a function of redshift ($z$) for the BINGO FRB mock sample. Furthermore, we use current datasets (Supernovae, Baryonic Acoustic Oscillations, and Cosmic Microwave Background data) prior to assessing the efficacy of constraining dark energy models using Monte Carlo methods.
Our results show that spatially localized BINGO FRB dataset will provide promising constraints on the population of host galaxies intrinsic DM and be able to measure the nuisance parameters present within a FRB cosmological analysis. They will also provide alternative estimates on other parameters such as the Hubble constant and the dark energy equation of state. In particular, we should see that BINGO FRB data can put constraints on the degenerate $w-H_0$ plane, which the CMB is incapable of measuring, allowing FRBs to be a viable alternative to BAO to constrain the dark energy equation of state.
We conclude that FRBs remain a promising future probe for cosmology and that the FRBs localized by the BINGO project will contribute significantly to our knowledge of the current cosmological model.

\end{abstract}



\section{Introduction}

Fast Radio Bursts (FRBs) were discovered 
while analyzing archival radio data from the Parkes radio telescope and reported by \cite{Lorimer:2007qn}. Since the first discovery of FRBs, a number of events have been observed by several radio telescopes over the world \citep{Thornton:2013iua,Spitler:2016dmz,Chatterjee:2017dqg,Tendulkar:2017vuq,Bochenek:2020zxn}. Such an important discovery led to a new field in astronomy and physics, namely the discovery, distribution, localization, and properties of these extremely energetic objects. From the point of view of astrophysics, the intrinsic behavior, properties, origin, and theoretical description of FRBs are of utmost importance \citep{Cordes:2019cmq,Petroff:2019tty,Petroff:2021wug}. 

FRBs are radio transients with milliseconds duration. Their distances are estimated from the dispersion measure (DM), which corresponds to the time distribution in frequency space \citep{Lorimer:2007qn,Dennison:2014vea,Caleb:2015uuk}. This is due to the passage of the wave train in the intergalactic medium, which is constituted of electrons, resulting in slightly different velocities according to the frequency as a consequence of electromagnetic interaction. The large DMs imply that the sources are at cosmological distance \citep{Keane:2012yh,Xu:2015wza,Chatterjee:2017dqg,Tendulkar:2017vuq}. 
Combined with the redshifts of their host galaxies, the DM of FRBs provides a new tool to investigate the structure of the Universe, including constraining the location of the ``missing baryons", probing the properties of the very diffuse intergalactic medium (IGM) and circumgalactic medium (CGM) around galaxies \citep{McQuinn:2013tmc,Ravi:2018ose,Macquart:2020lln,Bhandari:2021thi,Lin:2021syj}. 

The progenitors of FRBs are still a mystery to the astronomical community. Some theoretical models have been proposed to explain the origin of FRBs \citep{Platts:2018hiy,Zhang:2022uzl}; some possible progenitor candidates are compact object mergers \citep{Totani:2013lia,Wang:2016dgs}, the collapse of compact objects \citep{Falcke:2013xpa,Zhang:2013lta}, supernovae remnants \citep{Keane:2012yh,Connor:2015era,Cordes:2015fua}, and active galactic nuclei \citep{Romero:2015nec,Vieyro:2017flr}.
Utilizing the rapidly accumulating observational data of FRBs \citep{Petroff:2016tcr,CHIMEFRB:2021srp,Xu:2023did,CHIMEFRB:2023hfj}, the statistical properties of the key observed quantities, such as energy and waiting time, are extensively analyzed, yielding critical insights for investigating the origins and physical mechanisms of FRBs \citep{Law:2017rsz,Wang:2016lhy,Lin:2019ldn,Hashimoto:2022llm,Sang:2021cjq,Sang:2023zho,Wang:2022gmu,Sang:2024swg}.

It is also possible to apply FRBs surveys to the field of cosmology to study various critical issues \citep{Glowacki:2024cgu}, 
such as dark energy models \citep{Walters:2017afr,Zhao:2020ole,Qiu:2021cww,Zhang:2023gye,Zhao:2022bpd}, ionized gas fraction \citep{Yang:2016zbm,Jaroszynski:2018vgh,Walters:2019cie,Lin:2021syj}, spatial curvature density \citep{Yu:2017beg,Walters:2017afr}, cosmic baryon density \citep{Macquart:2020lln,Qiang:2020vta}, the baryon fraction in the IGM \citep{Lemos:2022kdh,Lemos:2023eaq}, Hubble parameter \citep{Wu:2020jmx,Zhang:2023gye}, as well as other cosmological questions \citep{Qiang:2019zrs,Walker:2018qmw,Zhu:2022mzv,Lin:2023jaq}.
The cosmological origin of FRBs is confirmed by identifying the host galaxy and its redshift measurement from other wavebands. The redshift is essential for probing the Universe using FRBs. If it is possible to locate the host of the FRB, one can foresee using the relation between the DM estimated from this object and the redshift of the host galaxy. This would permit further constraints in the cosmological parameters, which today is one of the paths towards understanding and improving our cosmological description, including the description of the Dark Sector of the Universe \citep{Abdalla:2020ypg}.

The aim of the BINGO (BAO from Integrated Neutral Gas Observations) telescope was primarily to inquire about the aforementioned constraints on cosmological parameters \citep{Abdalla:2021nyj,Abdalla:2021xpu,Costa:2021jsk,Wuensche:2021dcx}. With the further development of the ABDUS (Advanced Bingo Dark Universe Studies) as an extension to BINGO \citep{Abdalla2024}, it is clear that constraints on the FRBs parameters can become much more important and accurate, enabling a deeper view of the Universe details and a stronger constraint on the cosmological parameters as described above. 

When detected by a single-dish radio telescope, the FRB signal will probably be located within the primary beam of the instrument, posing an observational difficulty due to their usually large beams (e.g., the beam resolution of BINGO telescope is 40 arcmin). As a consequence, the host galaxy cannot always be accurately identified. However, localizations can likely be accurately produced in the radio band by performing an interferometric detection of a given FRB candidate in more than two radio dishes. Once the FRB is accurately localized, it is easy to match the FRB source with galaxy surveys and identify its host galaxy. 
So far, more than 110 FRBs have been localised, mainly from
the Commensal Real-time ASKAP Fast Transients (CRAFT) survey on the Australian Square Kilometre Array Pathfinder (ASKAP) telescope
\citep{Gordon:2023cgw,Shannon:2024pbu}, 
the Canadian Hydrogen Intensity Mapping Experiment Fast Radio Burst (CHIME/FRB) project \citep{Marcote:2020ljw,Kirsten:2021llv,Michilli:2022bbs,Bhardwaj:2023vha,Amiri:2025sbi}, 
the MeerKAT radio interferometer \citep{Rajwade:2022zkj,Caleb:2023atr,Driessen:2023lxj,Rajwade:2024ozu,Tian:2024ygd}, 
the Deep Synoptic Array (DSA-110) 
\citep{DeepSynopticArrayTeam:2023xmg,DeepSynopticArrayTeam:2023fxs,Connor:2024mjg,Faber:2024coo,Law:2023ibd,Sharma:2024fsq},
and the Karl G. Jansky Very Large Array (VLA) using the `realfast' fast transient detection system \citep{Chatterjee:2017dqg,Tendulkar:2017vuq,Law:2020cnm,Bhandari:2021pvj,Niu:2021bnl}.

With the increasing number of localized FRBs, an independent measurement of $H_0$ in the local Universe is expected to help resolve the Hubble tension.
\cite{Kalita:2024xae} use 64 localized FRBs to constrain $H_0$ with different models of IGM and host galaxy.
Similarly, \cite{Gao:2024kkx} constrained the Hubble constant with 69 localized FRBs. 
\cite{Wang:2025ugc} collected 92 localized FRBs and constrained the dark energy equation of state using a combination of Dark Energy Spectroscopic Instrument (DESI) baryon acoustic oscillation (BAO), PantheonPlus type Ia supernovae and cosmic microwave background (CMB) observations. 
\cite{Piratova-Moreno:2025cpc} built a catalog of 98 localized FRBs and employ three distinct methods to determine the most accurate value of the Hubble constant.
Recently, \cite{Acharya:2025ubt} use a sample of 110 localized FRBs to estimate the Hubble constant and the distribution of DM in the host galaxies.

The aim of this paper is to demonstrate the ability to constrain cosmology using FRBs localized by ABDUS \citep{Abdalla2024}. 
To achieve this goal, in a previous BINGO project IX paper, a BINGO/ABDUS Interferometry System (BIS) \footnote{In this work, we adopt the term `BINGO' to refer to the BINGO/ABDUS Interferometry System for convenience.} has been proposed to find the FRB sources in the sky via fringe fitting between the main telescope and the outriggers \citep{dosSantos:2023ipw}; hundreds of FRB detections and localizations are feasible. The simulation data produced in this previous work is used here as our mock for FRBs. With this framework, we will be able to show the cosmological advances that are possible with this dataset.

This paper is organized as follows: in Section \ref{sec:2}, we outline the methodology used in this paper; in Section \ref{sec:3}, we present the results of our simulations and how they improve current cosmological parameter constraints in current dataset; then we discuss the conclusions of our findings in Section \ref{sec:4}.

\section{Methodology and data}\label{sec:2}

\subsection{The Dispersion Measure}\label{sec:disp_meas}

FRBs are transient radio sources that commonly have extragalactic origin. The searches for FRBs are generally done by assembling many time series. When a FRB is detected, we observe a sharp pulse with dispersion effects at different frequencies. During the traveling from the FRB source to the telescope through an ionized medium, the radio photons interact with ionized matter, causing a time delay in the propagation of the pulse signal. The radio pulse is thus dispersed when traveling in the ionized intergalactic medium. The dispersion is quantified by the time delay of the pulse between the highest and lowest radio frequencies \citep{Petroff:2019tty},
\begin{equation}
    \Delta t  =\frac{e^2}{2\pi m_{\rm e} c} 
    \left(\nu_{\rm low}^{-2} - \nu_{\rm high}^{-2}\right) {\rm DM} \quad ,
\end{equation}
where $\Delta t$ is the arrival time delay in the observer frame,
$\nu_{\rm high}$ and $\nu_{\rm low}$ are the highest and lowest radio frequencies of observation, respectively.
Also, $m_{\rm e}$ is the mass of the electron, $c$ is the speed of light, and DM is the dispersion measure.
Thus, as the signal travels from the source and traverses the intervening medium to Earth, it directly traces ionized baryons that cannot be detected using other observational methods.

The DM is defined as the integral of the electron number density along the line of sight from the source to the observer,
\begin{equation}
    {\rm DM} = \int_0^d n_{\rm e}(l)dl,
    \label{eq:ne1}
\end{equation}
where $n_e$ is the free electron number density, $l$ is the path length and $d$ is the distance to the source. The DM is usually given in units of ${\rm pc/cm^3}$.

The main contributions to observed DM come from the Milky Way (MW) galaxy, intergalactic medium (IGM), and host galaxy (HG).
The total DM is the sum of these components:
\begin{equation}
{\rm DM}={\rm DM}_{\rm MW}+{\rm DM}_{\rm IGM}+\frac{{\rm DM}_{\rm HG}}{1+z},
\label{eq:DM_obs}
\end{equation}
where ${\rm DM}_{\rm MW}$ represents the amount of dispersion from the Milky Way.
It includes contributions of free electrons from the interstellar medium (ISM) and the Galactic halo.
The ISM part is derived from an electron distribution model, NE2001 \citep{Cordes:2002wz,Cordes:2003ik} or YMW16 \citep{Yao:2017kcp},
and the halo component has been estimated to be from tens to one hundred pc cm$^{-3}$ by numerous studies \citep{Prochaska:2019mnr,Keating:2020mnr,Macquart:2020lln,James:2021jbo,Cook:2023grs}.
${\rm DM_{IGM}}$ is the contribution from the intergalactic medium,
and ${\rm DM_{HG}}$ is the contribution from the host of the source itself.
Due to the expansion of the Universe, the observed DM for the host galaxy 
is scaled by a factor
of $1/(1+z)$.
The ${\rm DM_{HG}}$ strongly depends on the host galaxy type and its local environment \citep{Bassa:2017tke,Tendulkar:2017vuq}.
Following \cite{Deng:2013aga},
we defined the excess extragalactic contributions ${\rm DM_E}$ as
\begin{align}
{\rm DM}_{\rm E}
 &\equiv {\rm DM}-{\rm DM}_{\rm MW}
 = {\rm DM}_{\rm IGM}  + \frac{{\rm DM}_{\rm HG}}{1+z}.
\label{eq:DM_E}
\end{align}

The largest contribution to DM comes from the intergalactic medium, defined as 
\begin{equation}
    {\rm DM_{IGM}}= \int_{0}^{z_{\rm s}}\frac{c H_0 n_{\rm e}}{(1+z)^2 E(z)}dz, ~ ~
\end{equation}
where $H_0$ is the Hubble constant, $E(z)$ is the dimensionless expansion rate of the Universe at redshift $z$, and $z_{\rm s}$ is the redshift of the source. 

In the late Universe, the distribution of free electrons is highly inhomogeneous.
The electron number density and ${\rm DM_{IGM}}$ depend on the direction of the sky.
To find a relationship between DM and $z$, we assume the same approach as in \cite{Deng:2013aga}, where it was assumed that all baryons are homogeneously distributed and ionized with an ionization fraction $\chi_{\rm e}$. So in practice, we use the average of these values over the sky. The mean number density of free electrons at redshift $z$ is
\begin{align}
\bar{n}_{\rm e}(z)
 = \frac{3H_0^2\Omega_{\rm b}f_{\rm IGM}}{8\pi G m_{\rm p}}
\chi_{\rm e}(z)(1+z)^3,
\label{eq:ne}
\end{align}
where $f_{\rm IGM}$ is the fraction of baryon mass in the intergalactic medium and is often taken as 0.83 \citep{Deng:2013aga,Gao:2014iva,Yang:2016zbm}.  
The ionization fraction of free electrons $\chi_{\rm e}$ is 
\begin{equation}
  \chi_{\rm e}(z)=Y_{\rm H}\chi_{\rm e,H}(z)+\frac{1}{2}Y_{\rm He}\chi_{\rm e,He}(z), 
\end{equation}
where $Y_{\rm H}$ and $Y_{\rm He}$ are the hydrogen and helium mass fractions.
Their mass ratio is approximately $3:1$, normalized to $3/4$ and $1/4$, respectively.
For $z \lesssim 3$, the hydrogen and helium are fully ionized. So the ionization fractions are $\chi_{\rm e,H}=\chi_{\rm e,He}=1$.
Hence, the average of the DM from the intergalactic medium can be written as
\begin{align}
\left<{\rm DM}_{\rm IGM}(z_{\rm s})\right>  = \frac{3cH_0\Omega_{\rm b}f_{\rm IGM}}{8\pi G m_{\rm p}}
\int_{0}^{z_{\rm s}} \frac{\chi_{\rm e}(z)(1+z)}{E(z)}dz.
\label{eq:IGM}
\end{align}
Hence, Eq. (\ref{eq:IGM}) is a function of cosmological parameters. And if the redshift $z$ is measured, the DM can be used to constrain cosmological parameters. 

We assume ${\rm DM_{IGM}}$ follows a Gaussian distribution 
\begin{equation}
{\rm DM}_{\rm IGM}=N(\left<{\rm DM}_{\rm IGM}(z_{\rm s})\right>, \sigma_{\rm IGM}),
\end{equation}
where the average value of  ${\rm DM}_{\rm IGM}$ is expressed by Eq. (\ref{eq:IGM}).
The variance of ${\rm DM}_{\rm IGM}$ is typically assumed to be a constant in some previous literature. However, here we adopt a more appropriate and complex form where the variance of ${\rm DM_{IGM}}$ depends on the power spectrum \citep{McQuinn:2013tmc,Walker:2018qmw}:
\begin{align}
\sigma_{\rm IGM}^2\left({\rm DM},z_{\rm s} \right)
&= \int^{\chi_{\rm s}}_{0} d \chi (1+z)^{2}\bar{n}^{2}_{\rm e}(0)
\int\frac{d^2 k_{\perp}}{(2\pi)^{2}}P_{\rm e}(k_{\perp},z) \nonumber \\
&=\frac{c}{H_0}\left[\frac{\chi_{\rm e}\Omega_{\rm b}\rho_{\rm cr}}{\mu_{\rm e}m_{\rm p}}\right]^2
\int^{z_{\rm s}}_0 {\rm d} z \frac{(1+z)^2}{E(z)}\int {\rm d} k k P_{\rm e}(k,z),
\end{align}
where $z_{\rm s}$ is the redshift of the source, $\chi$ is the comoving distance, $\chi_{\rm s}$ is the comoving distance of the source, $\rho_{\rm cr}$ is the critical density and $\mu_{\rm e} \simeq 1.14$ is the mean mass per electron. $P_{\rm e}$ is the electron power-spectrum assumed to be $P_{\rm e}(k,z) = b\;P_{\rm m}(k,z)$, where $P_{\rm m}$ is the matter power-spectrum and $b\simeq1$ is a bias factor assumed to be constant.

Currently, there is still limited knowledge about the host term ${\rm DM}_{\rm HG}$ due to the lack of detailed observation on the local environment of most FRB sources.
It is difficult to accurately characterize ${\rm DM}_{\rm HG}$ because of its large uncertainty and the lack of well-localized FRBs.
It may vary in a large range, ranging from several tens to hundreds ${\rm pc/cm^3}$ \citep{Niu:2021bnl,Xu:2021qdn}.
To account for the large variation of ${\rm DM}_{\rm HG}$, its distribution is usually modeled by the log-normal function \citep{Macquart:2020lln,Zhang:2020mgq,Lin:2023yec}.
We adopt the probability distribution of ${\rm DM}_{\rm HG}$ as follows:
\begin{eqnarray}
P({\rm DM}_{\rm HG})= \frac{1} {\sqrt{2\pi} {\rm DM}_{\rm HG} \sigma_{\rm host}} \exp\left[- \frac{ \left(\ln {\rm DM}_{\rm HG} - \mu \right)^2 }{2 \sigma_{\rm host}^2}\right],
\end{eqnarray}
which is characterized by a median $\exp(\mu)$ and logarithmic width parameter $\sigma_{\rm host}$. 
Currently, there is no strong evidence for the redshift-dependence of $\mu$ or $\sigma_{\rm host}$ \citep{Tang:2023qbg}. 
Thus, we treat these two parameters as free constant parameters, although this could be modified when real data is being analyzed, as we are following a Bayesian approach.

\subsection{Cosmological models considered}

The dark energy equation of state (EoS) parameter is defined as $w(z)=p_{\rm de}(z)/\rho_{\rm de}(z)$, with pressure $p_{\rm de}(z)$ and density $\rho_{\rm de}(z)$ of dark energy, respectively. 
In a spatially flat Universe, the dimensionless Hubble expansion rate of the Universe at $z$ is given by the Friedmann equation:
\begin{align}
E(z)=\sqrt{
\Omega_{\rm m}(1+z)^{3}+\Omega_{\rm de}\exp \left[ 3\int_0^z \frac{1+w(z')}{1+z} d z' \right]}.
\end{align}
Here $E(z)=H(z)/H_0$. $H_0$ is the expansion rate measured today, and $\Omega_{\rm de}=1-\Omega_{\rm m}$ is the dimensionless dark energy density of the Universe, where we consider only flat models in this paper.
We assume three cosmological models outlined below:
\begin{enumerate}
\item the $\Lambda$CDM model, $w(z)= -1$, in which the cosmological constant $\Lambda$ serves as dark energy; \\
\item the $w$CDM model, $w(z)= w$, characterized by a constant EoS; \\
\item the $w_0w_a$CDM model, using the CPL parameterization \citep{Chevallier:2000qy,Linder:2002et}, $w(z)=w_0+w_az/(1+z)$, in which dark energy has an evolving EoS characterized by two parameters $w_0$ and $w_a$. 
\end{enumerate}

We choose the cosmological model of our mock catalog of FRB detection by the BINGO/ABDUS interferometric system to have the same cosmological parameters as the best-fit Planck cosmological fit. Therefore, we take $\Omega_{\rm b}=0.0494$, $\Omega_{\rm m}=0.3166$ and $H_0=67.27$ km/s/Mpc as the fiducial values from Planck team \citep{Planck:2018vyg}.

\subsection{Detailed FRB mock distribution}
In this work, we consider a spatial distribution $f_{\rm z}$ for the population of cosmological FRBs. Such distribution is assumed to be uniform in co-moving volume \citep{Luo:2018tiy,Luo:2020wfx,Chawla:2021igg,dosSantos:2023ipw}, 
\begin{equation}
f_z(z)\equiv\frac{\partial V}{\partial \Omega \partial z}
      =\frac{c}{1+z}\frac{r^2(z)}{H(z)}.
\label{eq:fz}
\end{equation}
However, this distribution alone cannot describe the population of observed FRBs at a given redshift. To do so, we need to take into account the particular characteristics of a specific survey. 

The interaction between the cosmological information and the telescope design is performed by the Python code FRBlip \citep{dosSantos:2023ipw} from the BINGO collaboration. In summary, the code generates the cosmological population, as Eq. (\ref{eq:fz}), in a given redshift range, associating each FRB to a luminosity distribution. Then FRBlip makes the interaction of the cosmological population with each of the telescope beams in auto and cross-correlations to calculate the signal-to-noise ratio (S/N) of each FRB in each baseline of the BIS, which is the telescope of interest here. More details, such as system temperature, gain parameters for the simulation, and S/N values to classify the FRBs, can be found in \cite{dosSantos:2023ipw}.

\begin{figure}[htbp]
\centering
\includegraphics[width=0.45\textwidth]{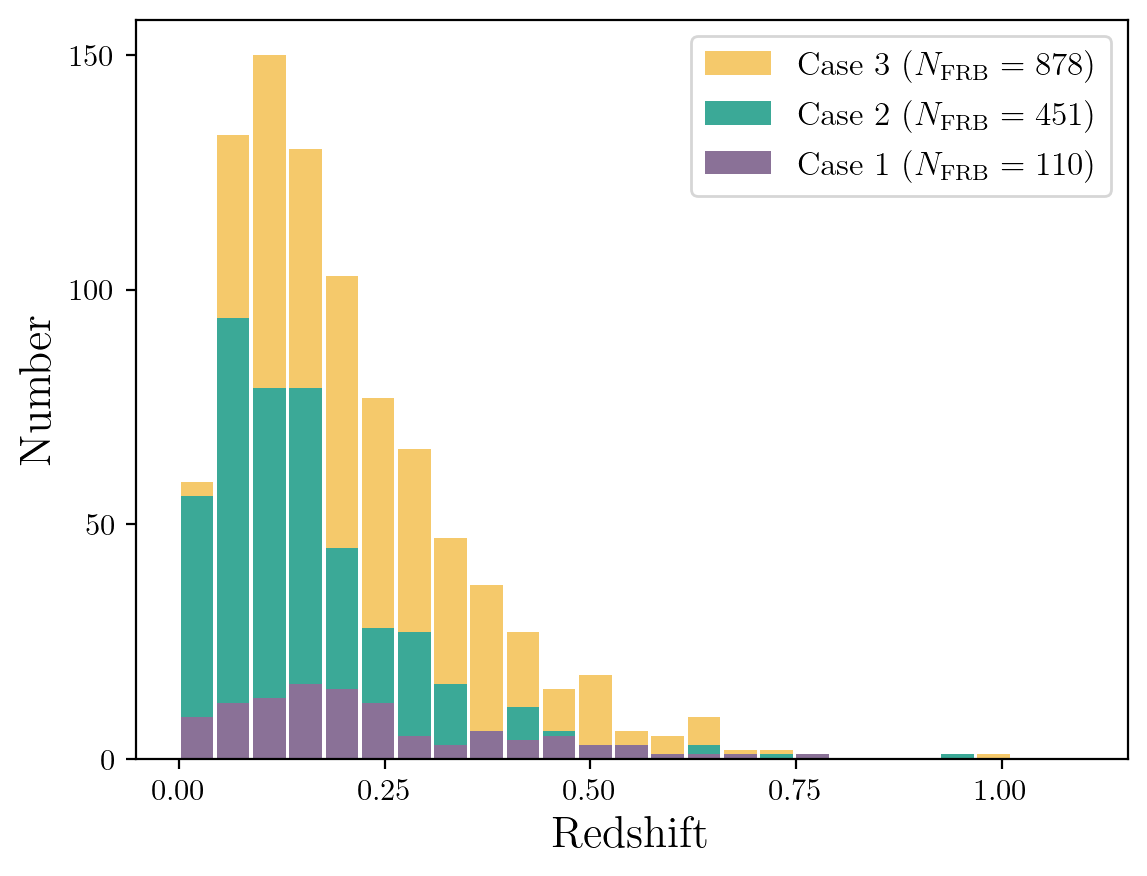}
\caption{\label{fig:z} The mock redshift distribution of the FRB localized by BINGO. The different colors represent different cases for a mock observation with different number of outriggers and outrigger configurations as described in the text, observed over 5 years. 
}
\end{figure}

\begin{figure*}[htbp]
\centering
\includegraphics[width=0.95\textwidth]{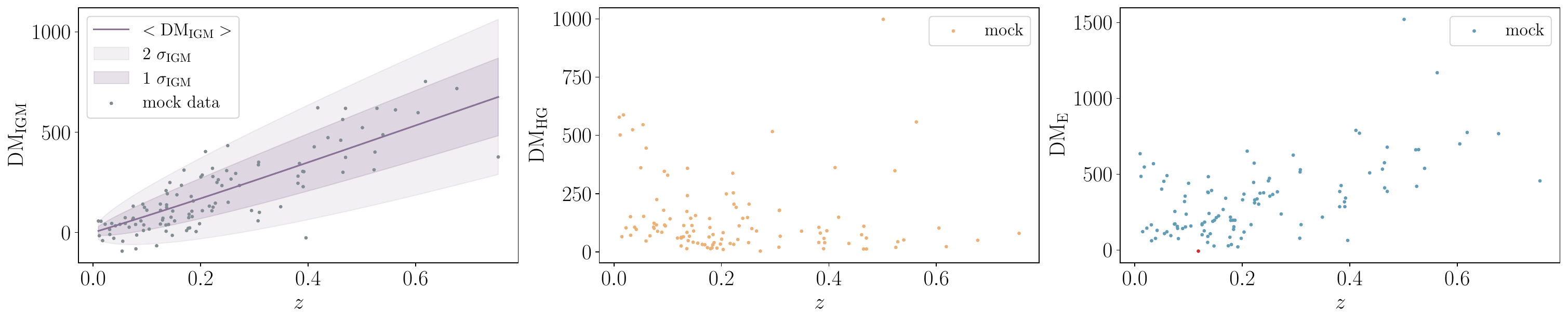}
\includegraphics[width=0.95\textwidth]{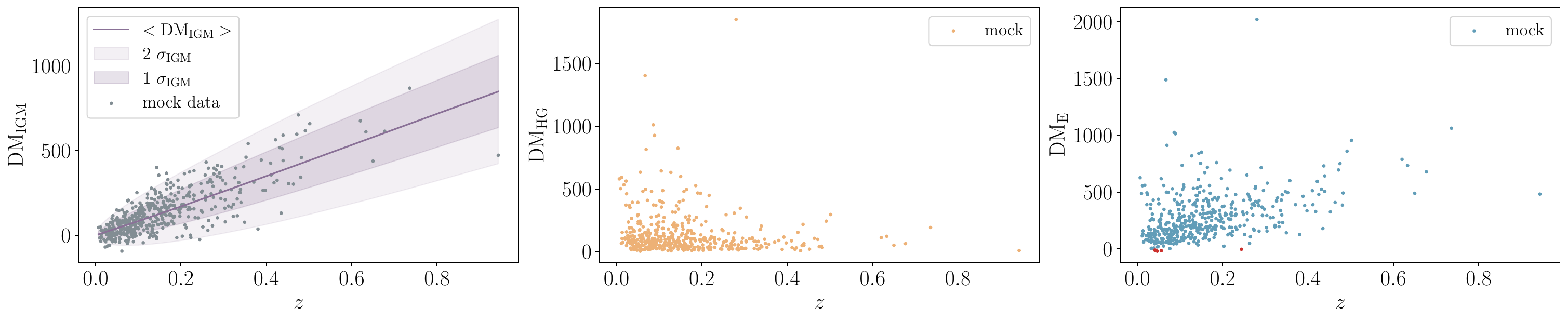}
\includegraphics[width=0.95\textwidth]{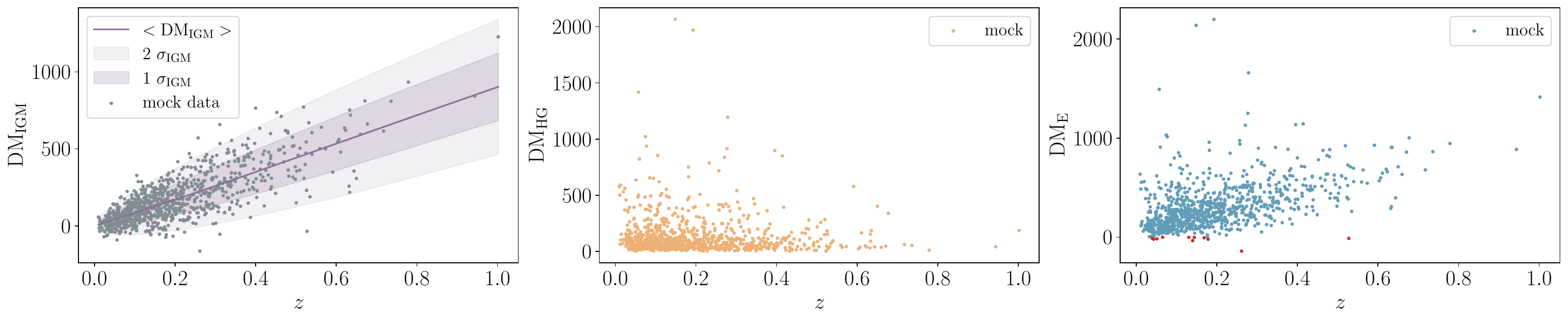}
\caption{\label{fig:hg} The simulated FRB data for a 5-year observation of BINGO.
The mock data for different cases corresponds to Case 1 ($N_{\rm FRB}=110$), Case 2 ($N_{\rm FRB}=451$), Case 3 ($N_{\rm FRB}=878$), respectively. 
The shadows represent the error intervals for the $1\sigma$ and $2\sigma$ confidence levels.
The red dots in the third column indicate that ${\rm DM_E}$ is negative, which is non-physical. 
Therefore, we removed these points from our later analysis. }
\end{figure*}

For this work, we use the same cosmological population as in \cite{dosSantos:2023ipw}, which was initially sampled in the redshift range of $0<z<10$. The mock FRBs used in our paper are selected from the cosmological population by considering the specific telescope parameters for 5 years of observation. We consider three different designs for ABDUS, varying the number of horns and outriggers. We plot in Figs. \ref{fig:z} and \ref{fig:hg} the number and the observed simulated distribution for FRBs that can be localized by ABDUS. Notably, as the number of stations and mirror size increases, the telescope can localize more FRBs, and their redshift will be higher. Below, we present the three scenarios we are considering in our analysis:
\begin{itemize}
      \item Case 1: BINGO Main station + outriggers with 6 m diameter collecting area capable of surveying 10 beams in the sky at different locations ($N_{\rm FRB}= 110$).  \\
      \item Case 2: BINGO Main station + outriggers with 4 m diameter collecting area capable of surveying 112 beams in the sky at different locations ($N_{\rm FRB}=451$). \\
      \item Case 3: BINGO Main station + outriggers with 6 m diameter collecting area capable of surveying 200 beams in the sky at different locations ($N_{\rm FRB}=878$).
\end{itemize}

For more details regarding Case 1 we refer the reader to a more detailled explanation of the simulation described in \citep{dosSantos:2023ipw}. For both Cases 2 and 3 the set up is similar to Case 1, however both these cases have a larger number of single dish antennas of 4m (Case 2) or 6m (Case 3) diameter. Cases 2 and Cases 3 can also be achieved by data collected by 3 phased array stations with multi-beam capabilities and with same diameters (4m and 6m). When considering the BINGO main telescope simulated data, all the possible auto and cross-correlations with its beams are considered in all cases. Only FRBs localized (with a detected signal to noise) in two or more baselines are used in this analysis, corresponding to the FRBs for which a reliable detection can be achieved.

\subsection{Likelihood}
Previous theoretical forecasts (e.g., \cite{Walters:2017afr}) has been done using the likelihood function of ${\rm DM}_{\rm E}$ as a Gaussian function, based on the assumption that ${\rm DM}_{\rm HG}$ follows a Gaussian distribution. As mentioned in Subsection \ref{sec:disp_meas}, 
different from \cite{Walters:2017afr}, 
we adopt the probability distribution of ${\rm DM}_{\rm HG}$ as a log-normal function 
which is characterized by a median $\exp(\mu)$ and logarithmic width parameter $\sigma_{\rm host}$. 
We treat these two parameters as free constant parameters. 
Therefore, in the likelihood function, we use a marginalization of the log-normal distribution.  Compared with \cite{Walters:2017afr}, our method is more conservative and has different results.

The probability density function (PDF) for ${\rm DM}_{\rm E}$ can be expressed as follows
\begin{align}
f({\rm DM_E}) =&\int_{0}^{{\rm DM_E}(1+z)}
\frac{1} {\sqrt{2\pi} \sigma_{\rm IGM}} \nonumber \\
&\times\exp\left[- \frac{({\rm DM_E} - \frac{{\rm DM_{HG}}}{1+z} - \left<{\rm DM}_{\rm IGM}\right> )^2 }{2\sigma_{\rm IGM}^2}\right] \nonumber \\
&\times\frac{1} {\sqrt{2\pi}{\sigma_{\rm host} \rm DM}_{\rm HG}}
\exp\left[- \frac{ \left(\ln {\rm DM_{HG}} - \mu \right)^2 }{2 \sigma_{\rm host}^2}\right]   d( {\rm DM_{HG}} ).
\end{align}
By setting $x=\ln {\rm DM_{HG}}$, the above equation changes to 
\begin{align}
f({\rm DM_E}) =&\int
\frac{1} {\sqrt{2\pi} \sigma_{\rm IGM}}
\exp\left[- \frac{({\rm DM_E} - {\rm e}^x /(1+z) - \left<{\rm DM}_{\rm IGM}\right> )^2 }{2\sigma_{\rm IGM}^2}\right] \nonumber\\
&\times\frac{1} {\sqrt{2\pi}{\sigma_{\rm host}}}
\exp\left[- \frac{ \left(x - \mu \right)^2 }{2 \sigma_{\rm host}^2}\right]   {\rm d} x.
\end{align}
We also tested the simplest Gaussian distribution case for ${\rm DM_{HG}}$, and the results are shown in the following Subsection \ref{sec:diff_like}.
Using the same method in  \cite{Macquart:2020lln},
we adopt the likelihood ratio test statistic:
\begin{align}\nonumber
{\cal D}(\Omega_{\rm m}, \Omega_{\rm b}, H_0, 10^9 A_s, n_s, \mu, \sigma_{\rm host})
= 2\log{\cal L}_{\rm max} \\ 
-2\log{\cal L}(\Omega_{\rm m}, \Omega_{\rm b}, H_0, 10^9 A_s, n_s, \mu, \sigma_{\rm host})
\end{align}
where ${\cal L}_{\rm max}$ is the maximum likelihood. 

To constrain the parameter spaces of the cosmological scenarios, 
we have used the publicly available Markov Chain Monte Carlo code CosmoMC \citep{Lewis:1999bs,Lewis:2002ah}\footnote{http://cosmologist.info/cosmomc/}, which
supports the Planck 2018 likelihood \citep{Planck:2019nip} and additionally has a convergence diagnostic following the Gelman-Rubin statistics \citep{Gelman:1992zz}.
GetDist \footnote{https://github.com/cmbant/getdist/} is also employed to plot the posterior distributions of the cosmological parameters.
We adopt the priors from Table \ref{tab:pa_in_as} for the parameters mentioned above.

\begin{table}[!htbp]
\caption{List of the fiducial values and prior ranges on the independent parameters using ``astro" parameterization in CosmoMC. In addition, we set $\Omega_{\rm b}H_0$ as a derived parameter.}
\centering
\begin{tabular}{l c c c }
  \hline
  Parameter & Fiducial value & Prior range
  \\\hline
  $\Omega_{\rm m}$ & 0.3166 & [0.01, 0.99]
  \\
  $\Omega_{\rm b}$ & 0.0494 & [0.01, 0.1]
  \\
  $H_0$ [km s$^{-1}$ Mpc$^{-1}$] & 67.27 & [20, 100]
  \\  
  $10^9 A_{\rm s}$ & 2.101 & [1.5, 2.5]
  \\
  $n_{\rm s}$ & 0.9649 & [0.8, 1.2]
  \\
  $e^{\mu}$ [pc/cm${^3}$] & 100 & [20, 200]
  \\
  $\sigma_{\rm host}$ & 1 & [0.2, 2]
  \\
  $w$ (i.e. $w_0)$ & -1 & [-3, 0]
  \\
  $w_a$ & 0 & [-3, 3]
  \\\hline
\end{tabular}
\label{tab:pa_in_as}
\end{table}

In our analysis, we note that it is possible to have a negative estimated value of the DM after host galaxy correction (and in a more realistic case after the Milky Way DM correction). This is simply due to statistical errors in the difference between the two quantities. This is largely ignored in this work, and we remove the values that, due to the errors, come across as negative values in our mock catalogs. This should have a minimal impact on the final results as the numbers of FRBs removed are small, around 1\% of the total FRBs. In a realistic scenario, we could use a more sophisticated likelihood to handle such cases, estimating the probability with which such negative values can arise. This approach would involve accounting for the error bars when comparing the extragalactic DM as well as the Milky Way's DM.

\subsection{Other observational data} \label{sec:other}
In this section, we describe the other observational datasets that have been used to constrain the underlying
cosmological models and the methodology of the statistical simulations. We produce a mix of simulated data combined with real data currently available for cosmological parameter estimation. It is perfectly possible to mix simulated data with real data in order to produce forecasts for future surveys. The observational datasets are there merely to provide constraints that are equivalent to the prior constraints that we would use in a real survey of FRBs. We argue here that this would be the most efficient way of producing such results simply because we know exactly what constraints we obtain from current data from other analyses of each of the collaborations. 

It is worth stressing that there is a concern associated with combining real data with simulations. Incompatibility between the datasets can lead to potential discrepancies. For example, it would be very difficult to simulate a mock of what FRBs would yield in terms of detection in the case of CMB data combined with $H_0$ local measurement data as both datasets are in tension with one another. We, therefore, avoid any such possibilities and remain within the cases where CMB data is compatible with the other datasets used. In other words, it is beyond of the scope of this analysis to identify sources of tension between cosmological datasets, however it is important to state that a FRB dataset such as the mock we are analysing can potentially help in solving these tensions. Below, we describe the datasets used in the next section of our analysis.

We use the following observational datasets in the analysis:
\begin{itemize}
  \item Cosmic Microwave Background (CMB): in this work, we utilize the CMB data from the final release by the Planck collaboration. Specifically, we have employed the CMB temperature and polarization power spectra, which include TTTEEE+lowl+lowE provided by Planck 2018 \citep{Planck:2018vyg}.
  \item Baryon Acoustic Oscillations (BAO): we performed a joint analysis of BAO using various datasets in this work. We specifically combined measurements from the 6dFGS \citep{Beutler:2011hx}, SDSS MGS \citep{Ross:2014qpa}, and BOSS DR12 \citep{BOSS:2016wmc} as referenced by the Planck 2018 team.
  \item Supernovae Type Ia (SN Ia): 
  in this study, we utilize a recent dataset of SNIa known as the Pantheon sample \citep{Pan-STARRS1:2017jku}, which is distributed across a redshift range of  $z\in[0.01, 2.3]$.
\end{itemize}

\section{Results}\label{sec:3}

\begin{figure*}[htbp]
\centering
\includegraphics[width=0.95\textwidth]{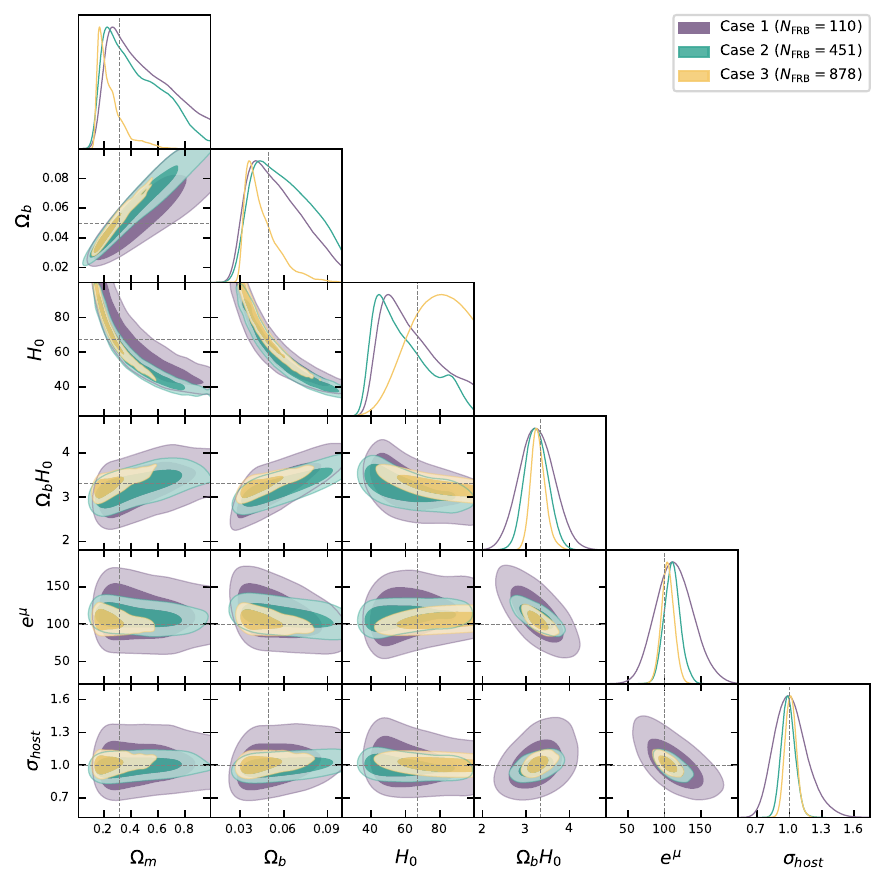}
\caption{\label{fig:frb_lcdm} 
Constraints of cosmological parameters in $\Lambda$CDM model from various scenarios described in the text for BINGO varying the number of horns. 
The purple contour, green contour and yellow contour correspond to 
Case 1 ($N_{\rm FRB}=110$), Case 2 ($N_{\rm FRB}=451$) and Case 3 ($N_{\rm FRB}=878$), respectively. 
The unit of $H_0$ is km s$^{-1}$ Mpc$^{-1}$, and the units of $e^\mu$ is pc/cm${^3}$.
The dashed line stands for the fiducial values we use in this work.}
\end{figure*}

In this section, we simulate the FRB observations from BINGO in order to constrain the cosmological parameters. 
The prior ranges of free parameters are listed in Table \ref{tab:pa_in_as}. 
In addition, we shall compare and combine these observations with data from CMB, BAO, and SN Ia datasets. It is worth noting that these constraints are based on observational data, which are used as prior information for our analysis.  This approach is valid when there are no significant tensions in the parameters. 
We will also discuss the constraints on dark energy parameters in $\Lambda$CDM, $w$CDM, and $w_0w_a$CDM models using our FRB observations.

\subsection{Constrains from BINGO FRB alone}

\begin{figure*}[htbp]
\centering
\includegraphics[width=0.95\textwidth]{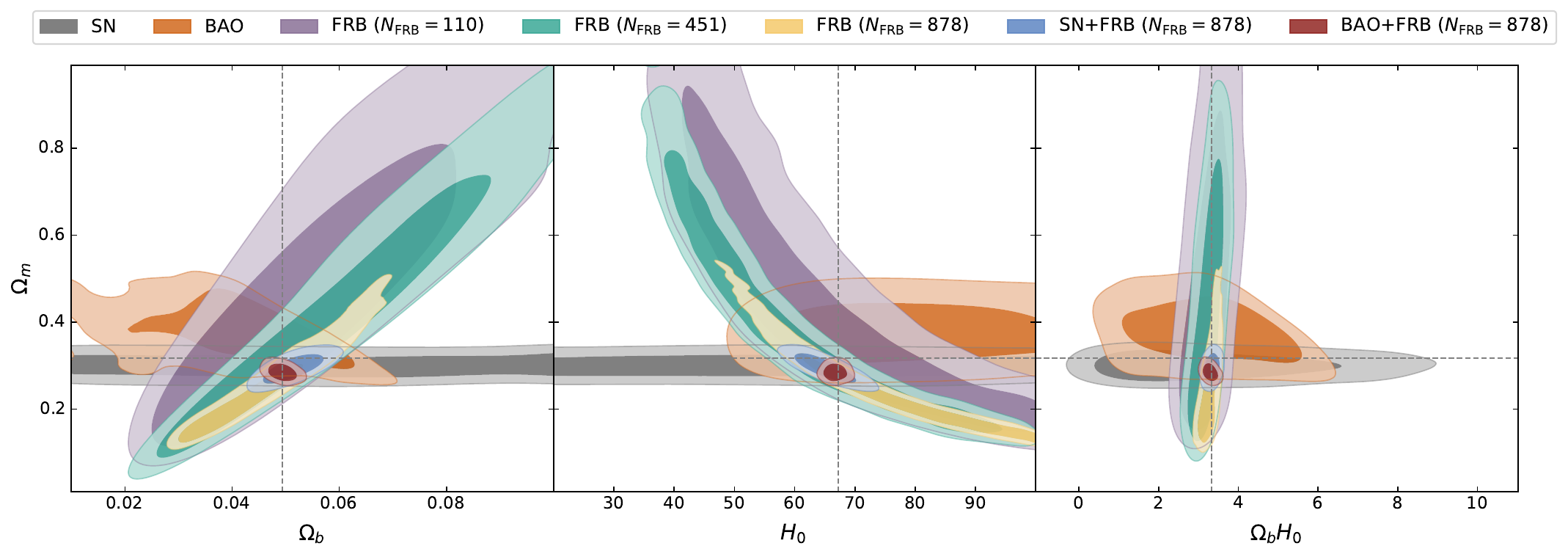}
\caption{\label{fig:frb_sn_bao_lcdm} Constraints ($1\sigma$ and $2\sigma$ confidence level) on
$\Omega_{\rm m}$, $\Omega_{\rm b}$, $H_0$ and $\Omega_{\rm b} H_0$ for $\Lambda$CDM model using the SN, BAO, FRB and their combined dataset. 
The dashed lines in the three panels indicate the fiducial value chosen for the FRB simulated data, which is consistent with and the same as the Planck best-fit values for $\Lambda$CDM parameters.}
\end{figure*}

\begin{figure*}[htbp]
\centering
\includegraphics[width=0.68\textwidth]{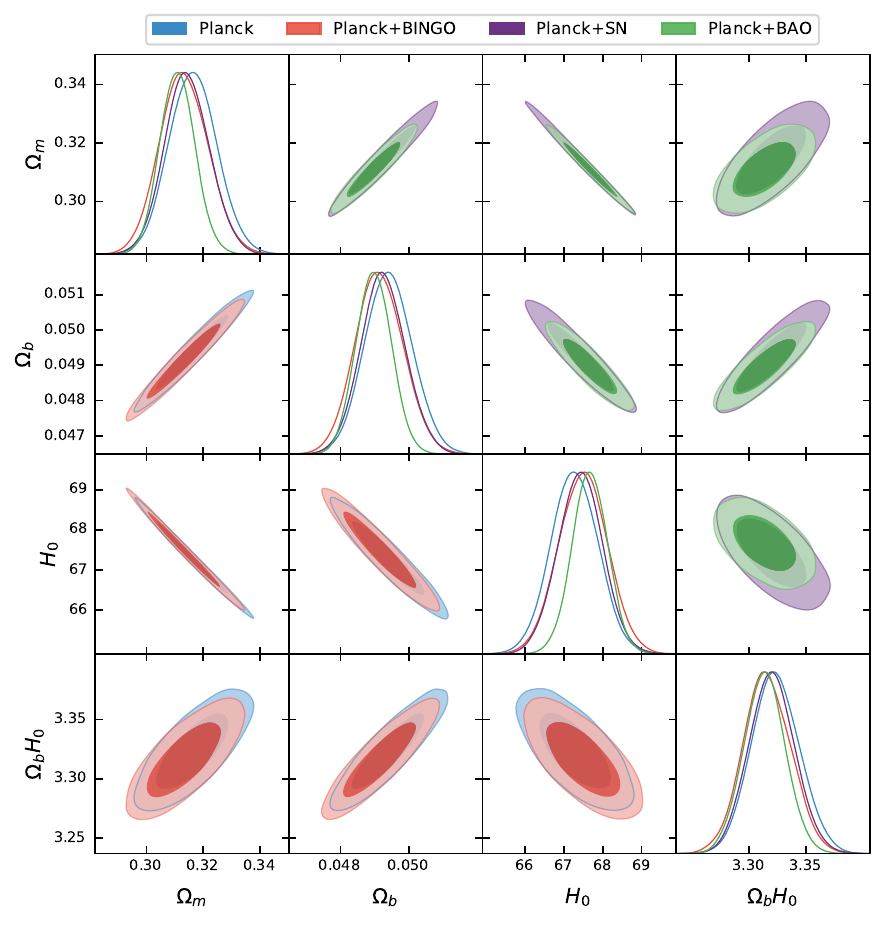}
\caption{\label{fig:plc_frb_sn_bao_lcdm} Constraints ($1\sigma$ and $2\sigma$ confidence level) on
$\Omega_{\rm m}$, $\Omega_{\rm b}$, $H_0$ (in unit of km s$^{-1}$ Mpc$^{-1}$) and $\Omega_{\rm b} H_0$ for $\Lambda$CDM model using the CMB, CMB+FRB, CMB+SN and CMB+BAO data. Here ``BINGO'' denote the Case 3 ($N_{\rm FRB}=878$)}.
\end{figure*}

\begin{table*}[htbp]
    \caption{68\% constraints on the free and derived parameters of the $\Lambda$CDM model using several observational datasets.}
    \centering
    \begin{tabular}{c c c c c c c}
        \hline
        Parameters &  Planck & Planck+BINGO & Planck+SN & Planck+BAO & SN+BINGO & BAO+BINGO 
        \\\hline
        $\Omega_{\rm m}$ & 
        $0.3165^{+0.0083}_{-0.0084}$ & $0.3132^{+0.0082}_{-0.0084}$ & 
        $0.3144^{+0.0076}_{-0.0081}$ & $0.3110^{+0.0061}_{-0.0062}$ & 
        $0.2960^{+0.0214}_{-0.0219}$ & $0.2870^{+0.0118}_{-0.0168}$
        \\
        $\Omega_{\rm b}$ & 
        $0.04940\pm0.00069$ & $0.04913^{+0.00067}_{-0.00068}$ & 
        $0.04925\pm0.00065$ & $0.04897^{+0.00051}_{-0.00052}$ &
        $0.05130^{+0.00326}_{-0.00362}$ & $0.04933^{+0.00168}_{-0.00170}$
        \\
        $H_0$ [km s$^{-1}$ Mpc$^{-1}$] & 
        $67.28^{+0.61}_{-0.60}$ & $67.51^{+0.62}_{-0.60}$ & 
        $67.43^{+0.56}_{-0.57}$ & $67.67^{+0.46}_{-0.45}$ &
        $65.07^{+2.95}_{-3.17}$ & $66.78^{+1.16}_{-1.26}$
        \\
        $\Omega_{\rm b} H_0$ & 
        $3.323\pm0.021$ & $3.316\pm0.020$ & 
        $3.320\pm0.020$ & $3.314\pm0.018$ &
        $3.329^{+0.127}_{-0.126}$ & $3.294^{+0.115}_{-0.117}$
        \\\hline
    \end{tabular}
    \label{tab:plc_frb_sn_bao}
\end{table*}

In this subsection, we discuss the potential of simulated BINGO FRB data to constrain the aforementioned parameters. First, we emphasize that FRB data will have significant degeneracy alone. If we look at Eq. (\ref{eq:IGM}) that describes the dispersion measure of FRBs for the IGM as a function of redshift, we can clearly see that there is a perfect degeneracy between the baryon fraction, the Hubble constant, and the fraction of gas contained within the IGM. In our analysis, we have chosen $f_{\rm IGM}$ to be constant, as it is perfectly degenerate with $H_0$ and $\Omega_{\rm b}$. This is one reason why we choose to express our constraints as the product of $H_0\times\Omega_{\rm b}$ in some of our graphs. This shows the power of constraint for FRBs and shows that if FRB cosmology will be able to put constraints on the Hubble constant, it will do so within a factor that depends on our knowledge of $f_{\rm IGM}$.

Having pointed out the degeneracy issue, we present the FRB-only constraints within the $\Lambda$CDM model in Fig. \ref{fig:frb_lcdm}. We note that the chosen parameters $\mu$ and $\sigma_{\rm host}$ are nuisance parameters with respect to the modeling of the FRB population distribution. We clearly see that both parameters are reasonably constrained for all simulated BINGO setups. This suggests that, in terms of the statistical power of internally calibrating the mean value for the ${\rm DM_{HG}}$ within the host galaxies, as well as its associated scatter, our model presents reliable results.

We can see that there is a good constraint on the product of $\Omega_{\rm b} \times H_0$ and that for enough FRB detections, there is a reasonable constraint on $\Omega_{\rm m}$ for a $\Lambda$CDM cosmology. This suggests that, when combined with  cosmological probes mentioned in Sec. \ref{sec:other}, FRB data has the potential of helping to put constraints in different cosmological models.

It is important to note that the Planck results are used as fiducial values in our analysis. Given the current tension between CMB measurements and local cosmological measurements of the Hubble constant, it is possible that the Planck-derived value may be incorrect, while studies such as SH0ES may provide a more reliable estimate with a higher $H_0$ measurement. Consequently, the uncertainties associated with the parameter estimates in our results are of greater significance than the central values themselves.

The capability of the BINGO telescope to detect and localize FRBs was investigated in our previous study \citep{dosSantos:2023ipw}. Furthermore, the next phase following BINGO has been planned as the ABDUS project \citep{Abdalla2024}. In order to improve cosmological constraints, a larger number of FRBs need to be localized, which could be achieved by increasing the number of outriggers, enlarging the size of their mirrors, or employing phased-array and multi-beam technologies. For example, ABDUS is expected to localize approximately 20 events per year when using 10 outriggers with 4 m mirrors and two baselines. In contrast, if 50 sets of three outriggers (a total of 150 outriggers) are directed toward the same sky region, the number of localized FRBs increases to approximately 180 events per year, based on localization by at least three baselines.

\subsection{Comparison with BAO, SNIa, and CMB}

We now turn our attention to studying how the FRB data can enhance or constrain the current cosmological $\Lambda$CDM model given the cosmological datasets mentioned in Sec. \ref{sec:other}. 
We plot our results in Figs. \ref{fig:frb_sn_bao_lcdm} and \ref{fig:plc_frb_sn_bao_lcdm}, and Table \ref{tab:plc_frb_sn_bao} depicts the best fit of cosmological parameters in our investigation. 

In Fig. \ref{fig:frb_sn_bao_lcdm}, we can see that there are degeneracy constraints that are significantly orthogonal in the BINGO, BAO, and SNIa datasets when looking at the $\Omega_{\rm m}$ versus $\Omega_{\rm b} H_0$ plane. This means that the combination of these data can potentially yield interesting information to confirm the data constraints from Planck even within a $\Lambda$CDM scenario. In the central panel ($H_0 \times \Omega_{\rm m}$) that a combination of the distance measures arising from FRB, SNIa, and BAO can yield an interesting and novel constraint on the Hubble constant. This might be of interest, especially given the Hubble tension that exists in the current cosmological data. It is important to note that these constraints rely on choosing the fiducial value of $f_{\rm IGM}$, with an upper limit $f_{\rm IGM}=1$. This means that, even considering this astrophysical nuisance parameter, we should be able to add information to investigate the Hubble tension with a future combination of distance measures, including FRB data. 

Notably, the red and blue contours suggest that the BINGO FRB constraints are highly complementary to the SNIa and BAO constraints. We note that the BAO + BINGO constraint is somewhat lower than the nominal simulated value for the matter density. This is possible because we have used the Planck central values as the fiducial values for the BAO constraints, rather than the combined Planck + BAO central value. The discrepancy is small enough, and within the error bars in Table \ref{tab:plc_frb_sn_bao}, and we are not concerned about it in this work.

In Fig. \ref{fig:plc_frb_sn_bao_lcdm}, we notice that, for the $\Lambda$CDM model, the quality of the Planck data is so high that the other cosmological probes used in this work add very little to the cosmological information brought by Planck only. FRB data marginally improve the cosmological contours, and it is remarkable in the case of Planck combined with BINGO FRB simulated datasets. BAO and SNIa Pantheon data also minimally contribute to improvement of cosmological constraints, as can be seen from the data in Table \ref{tab:plc_frb_sn_bao}. 

\subsection{$w$CDM and $w_0w_a$CDM model}

\begin{figure}[!htbp]
\centering
\includegraphics[width=0.48\textwidth]{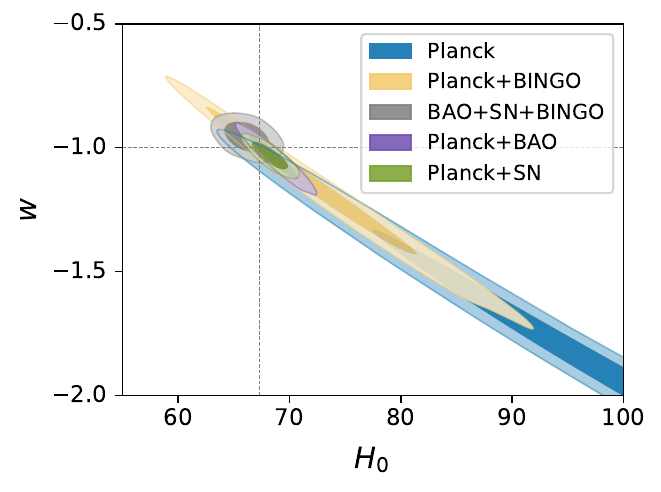}
\caption{\label{fig:wcdm} Comparison between Planck, Planck+BINGO, BAO+SN+BINGO, Planck+BAO and Planck+SN coinstraints in a $w$CDM model.
The dashed line indicates $w=-1$, which corresponds to the standard $\Lambda$CDM model. We note that the BINGO FRB mock constraints are able to constrain a combination of the $w$ and $H_0$ (in unit of km s$^{-1}$ Mpc$^{-1}$) without using any CMB data, which indicates this will be a powerful dataset to help clarify any Hubble tension in future datasets.
}
\end{figure}

\begin{figure*}[ht]
\centering
\includegraphics[width=0.45\textwidth]{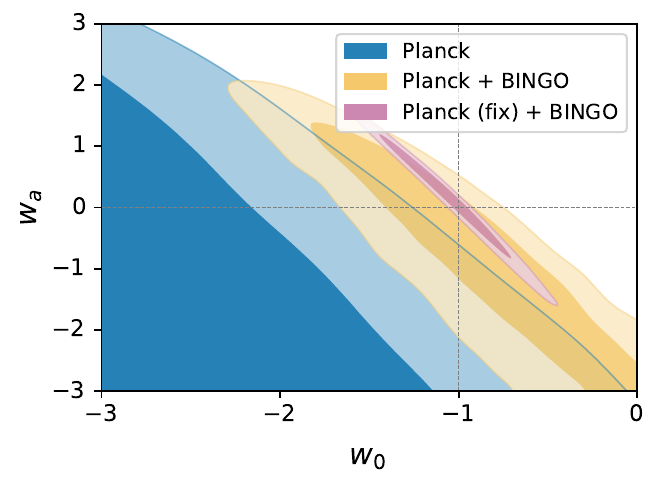}
\includegraphics[width=0.4\textwidth]{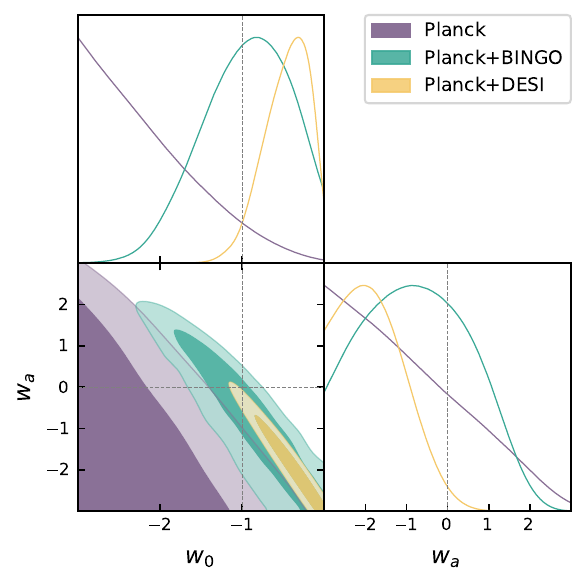}
\caption{\label{fig:cpl} 
Marginalized posterior distributions of the $(w_0, w_a)$ parameters are compared for various data combinations in the CPL model, including Planck-only, Planck+BINGO, and Planck (fix)+BINGO. Planck (fix) refers to fixing the cosmological parameters derived by Planck and only setting $w_0$ and $w_a$ free. Also, the posterior distribution and correlation contour plot of parameters $w_0$ and $w_a$ in the $w_0 w_a$CDM model are generated using datasets from Planck, Planck+BINGO, and Planck+DESI.  }
\end{figure*}

In the following, we discuss the constraints on the EoS of dark energy by using the FRB data. 
In Figs. \ref{fig:wcdm} and \ref{fig:cpl}, we show our constraint results on dark-energy EoS parameters for the $w$CDM and $w_0w_a$CDM models, respectively.

For the $w$CDM model, Planck CMB measurements cannot constrain the EoS parameter of dark energy.
By combining the CMB data with the simulated FRB samples, we note that, whereas the $\Lambda$CDM model for Planck is constrained in a very accurate way, this is not true for the case of $w$CDM model, where there is a long degeneracy for various values of $w$. Furthermore, it is shown that this long degeneracy can be broken by the measurements of one of the other cosmological parameters, which is why BAO or SNIa do manage, in conjunction with the CMB, to put constraints on the equation of state. Our results show that the BINGO FRB constraints can perform in a similar light to the BAO or SNIa constraints. In our results, shown in Fig. \ref{fig:wcdm}, we can see that the degeneracy from the Planck results is cut by the inclusion of the BINGO FRB dataset. Therefore, we can conclude that the results show good prospects for this, but it is possible that another tension might shed some light on the Hubble tension. 

The results are similar but somewhat more complex for the case of $w_0w_a$CDM model. As we know, the recent DESI results show a preference for a Universe where the equation of state parameter varies with redshift. For the $w_0w_a$CDM model, the CMB data alone can not provide a result of ($w_0$,$w_a$) and the best fit for Planck data even falls away from the $\Lambda$CDM scenario. These results need to be confirmed but it is possible that another tension might be appearing in our cosmological model. We present the results demonstrating how BINGO could complement CMB data and compare it with the findings of DESI added to the same Planck data. We can see from the left plot in Fig. \ref{fig:cpl} that the BINGO FRB results will be able to start casting light on the confirmation of the constraints for the CPL parameterization of the dark energy.

The present research also aims at studying the deep structure of the Dark Sector, especially a possibility of interacting dark energy and dark matter, implying a possible nontrivial extension of the elementary particles constitution to contain the Dark Sector \citep{Wang:2016lxa,Wang:2024vmw}. The possibility of a nontrivial dark energy equation of state is a step in that direction.

\subsection{Effect of alternate assumptions: the distribution of ${\rm DM_{HG}}$ and $\sigma_{\rm IGM}$}\label{sec:diff_like}

It is worth underscoring that the likelihood function for the BINGO FRB simulated data is highly dependent on some astrophysical distributions and properties of the respective FRBs. One important point to note is that the likelihood function incorporates the distribution of the mean ${\rm DM_{HG}}$, stemming from galaxy evolution and formation. In a real analysis, we would attempt to model this distribution and choose, maybe with the aid of a statistical tool such as the evidence, the best model that fits the data. However, in this analysis, we simply simulate ${\rm DM_{HG}}$ as the simplest Gaussian distribution to test how cosmological parameters are affected by different likelihoods. 

\begin{figure}[htbp]
\centering
\includegraphics[width=0.43\textwidth]{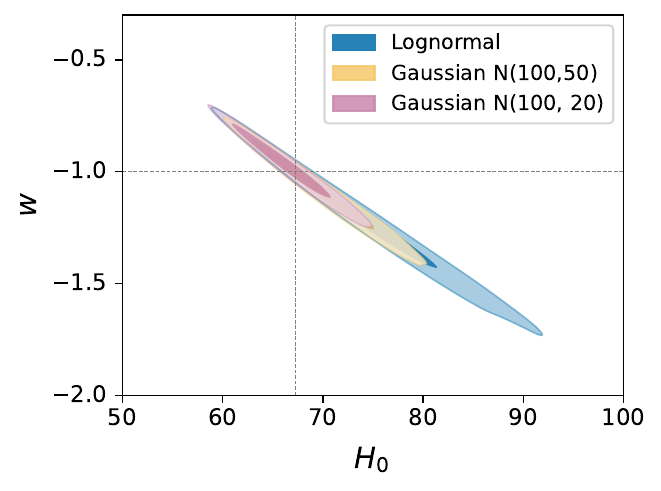}
\includegraphics[width=0.41\textwidth]{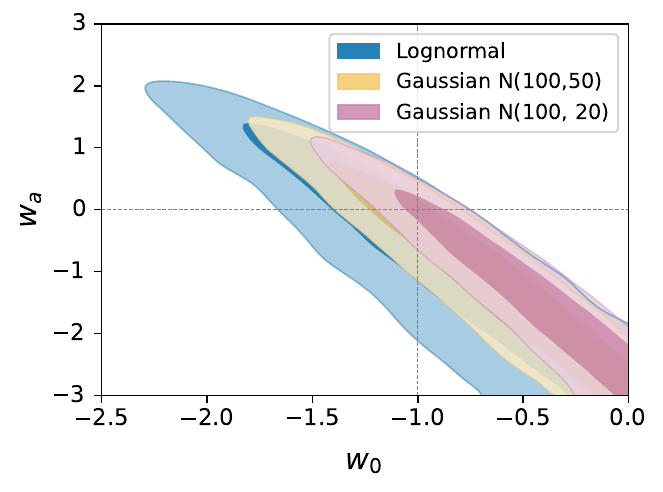}
\caption{\label{fig:lognormal} Compare Lognormal with Gaussian distribution of ${\rm DM_{HG}}$ in $w$CDM model. Also, marginalized posterior distributions of the $(w_0, w_a)$ parameters are compared for various distribution of ${\rm DM_{HG}}$ in $w_0 w_a$ model, including Lognormal, Gaussian $N(100, 50)$ and Gaussian $N(100, 20)$. The unit of $H_0$ is km s$^{-1}$ Mpc$^{-1}$.}
\end{figure}

We consider two different Gaussian distributions, namely, $N(100, 50)$ and $N(100, 20)$. We compare in Fig. \ref{fig:lognormal} the respective contours produced in previous sections, changing only the likelihood, which relates to the assumed distribution of the host galaxy DM. As we can see, the distribution does affect the contours as expected. Also, we find that the lognormal case, which has been used in the main body of the paper, is less well-constraining than assuming that the distribution from Gaussian ones. Presumably, because the log-normal distribution has a less compact distribution of values as opposed to the Gaussian distribution, it would have fewer outliers in terms of high DMs arising from the host galaxy. This more compact distribution presumably produces less degeneracy with the effect that the cosmological parameters have on the likelihood of degeneracy we have with these parameters. More specifically, these roughly halve the constraints we find on the $w-H_0$ plane, as we can see from the top panel of the figure. This means that if the real distributions of DM from host galaxies are more compact than an assumed log-normal distribution, then the constraints provided by a BINGO FRB survey could be significantly better than the constraints yielded in this analysis, and also the ability to measure the host galaxy contribution to the total DM budget could be improved.

\begin{figure}[htbp]
\centering
\includegraphics[width=0.45\textwidth]{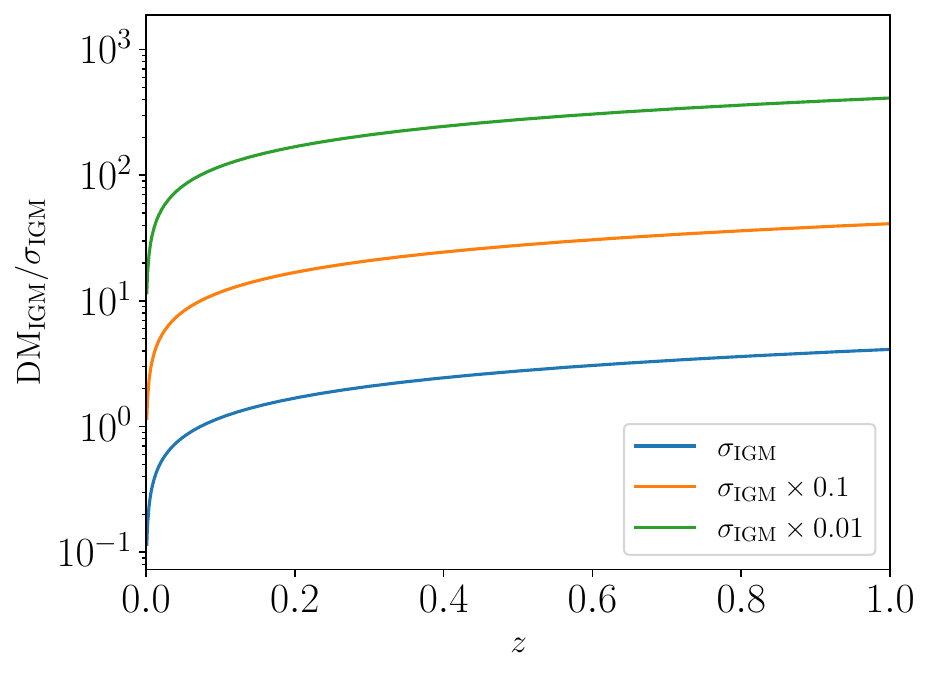}
\caption{\label{fig:dmsdm} The evolution of ${\rm DM_{IGM}}/\sigma_{\rm IGM}$ with redshift $z$ according to different bias factor $b$. }
\label{fig:dmsigmaigm}
\end{figure}

\begin{figure}[!htbp]
\centering
\includegraphics[width=0.48\textwidth]{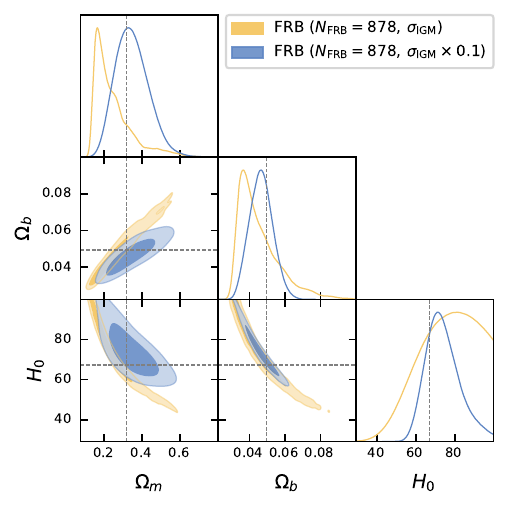}
\caption{\label{fig:sigma} Comparison of the result using  $\sigma_{\rm IGM}$ and $0.1\sigma_{\rm IGM}$, namely using a different bias factor $b$. The unit of $H_0$ is km s$^{-1}$ Mpc$^{-1}$.}
\label{fig:sigmaigm}
\end{figure}

\cite{Takahashi:2020wke} measured the free-electron power spectrum from hydrodynamic simulations and found the bias is unity at large scales ($k< 1{\rm Mpc^{-1}}$) but is strongly suppressed at small scales ($k> 1{\rm Mpc^{-1}}$). We have assumed the free-electron bias $b\simeq1$, which overestimates the $\sigma_{\rm IGM}$.
Here we investigate the effect of the bias parameter on the parameter constraint. We can see from Fig. \ref{fig:dmsigmaigm} what effect the bias parameters, which effectively modulates the value of the actual $\sigma_{\rm IGM}$, has in the parameter constraint. Given that the bias parameter modifies how much the DM follows the distribution of dark matter in the Universe, a change in bias is equivalent to having a different power spectrum with changed amplitude for the distribution of free electrons in our Universe and therefore, a larger bias implies a larger scatter of the actual values of the DM measured in the FBR mock data. We can see the equivalent effect in Fig. \ref{fig:sigmaigm}. A smaller value for the bias implies on tighter constraints, which we can see from the parameter degeneracy between $\Omega_{\rm b}$ and $H_0$. It is worth noting that according to some analysis, the value of this bias should indeed be less than one \citep{Shaw:2011sy,Takahashi:2020wke}.

\section{Conclusion}
\label{sec:4}
In this paper, we have taken simulated FRB mock datasets that we expect to have with a BINGO survey. We have assumed such FRBs are located in the sky via the Bingo Interferometric System (BIS) to a good enough accuracy that the host galaxy and, therefore, the host redshift can be identified in the catalog. This in itself is a great effort that would have to be done by the collaboration in terms of both detection and follow-up of the survey.

We have shown that depending on the number of outriggers and technology of the outriggers used in different scenarios, an expected number of sources detected can range from a hundred to almost a thousand FBRs with a full survey, we further show that these sources can provide us with significant astrophysical and cosmological information.

A survey of even a hundred FRBs, such as the most conservative estimate for the BIS, can yield a self-consistent measurement of the astrophysical properties of the DM of galaxies and its scatter. This should be of great importance to galaxy formation as it would unveil the physics behind the content of electrons, the free electrons density within the host galaxies at cosmological redshift. 

We also find that the FRBs from a BIS can complement the CMB data in the same way that SNIa and BAO do. Therefore, with good enough data, we can help shed light on interesting problems within the cosmological model today, such as the actual evolution of the dark energy equation of state, and we can also help shed light on the actual value of the Hubble constant. 

We conclude that the data provided by a BIS can greatly help us further understand the mysteries of the Universe and help our understanding of galaxy formation as well as the nature of the components of our Universe.

\begin{acknowledgements}
The BINGO project is supported by FAPESQ-PB, the State of Paraiba, FINEP, and FAPESP-SP, Brazil, 
by means of several grants. 
X.Z. is supported by grant from NSFC (Grant No. 12005183). 
Y.S. is supported by grant from NSFC (Grant No. 12005184, 12175192). 
G.A.H. is supported by Dean's Doctoral Scholarship (DDS) and Coordination for the Improvement of Higher Education Personnel (CAPES).
The authors acknowledge the National Laboratory for Scientific Computing (LNCC/MCTI, Brazil) for providing HPC resources of the SDumont supercomputer, which have contributed to the research results reported within this paper. URL: http://sdumont.lncc.br. 
F.B.A thanks the University of Science and Technology of China and the Chinese Academy of Science for grant number KY2030000215.
E.A. is supported by a CNPq grant.
A.R.Q. work is supported by FAPESQ-PB. A.R.Q. acknowledges support by CNPq under process number 310533/2022-8. 
A.A.C acknowledges financial support from the National Natural Science Foundation of China (grant 12175192) and Conselho Nacional de Desenvolvimento Científico e Tecnológico (CNPq) -Brazil (grant 102734/2024-0).
C.A. Wuensche thanks CNPq for grants 407446/2021-4 and 312505/2022-1, the Brazilian Ministry of Science, Technology and Innovation (MCTI) and the Brazilian Space Agency (AEB) who supported the present work under the PO 20VB.0009.
L.X. is supported by the National Research Foundation of Korea (NRF) through grant No. 2020R1A2C1005655 funded by the Korean Ministry of Education, Science and Technology (MoEST).
\end{acknowledgements}


\bibliographystyle{aasjournalv7}
\bibliography{references}



\end{document}